# Differential Analysis of Pseudo Haptic Feedback: Novel Comparative Study of Visual and Auditory Cue Integration for Psychophysical Evaluation

Nishant Gautam[1,2], Somya Sharma[1], Peter Corcoran[2], Kaspar Althoefer[1]

[1]Advanced Robotics @ Queen Mary University of London, London E1 4NS, United Kingdom
[2]Centre of Computational Cognitive and Connected Imaging (C3I),University of Galway

`nishant.gautam@universityofgalway.ie`[1,2]`, somyasharma08@gmail.com`[1]`,`
`peter.corcoran@universitofgalway.ie`[2]`, k.althoefer@qmul.ac.uk`[1]

**Abstract.** Pseudo-haptics exploit carefully crafted visual or auditory cues to trick the brain into "feeling" forces that are never physically applied, offering a low-cost alternative to traditional haptic hardware. Here, we present a comparative psychophysical study that quantifies how visual and auditory stimuli combine to evoke pseudo-haptic pressure sensations on a commodity tablet. Using a Unity-based *Rollball* game, participants (n = 4) guided a virtual ball across three textured terrains while their finger forces were captured in real time with a Robotous RFT40 force–torque sensor. Each terrain was paired with a distinct rolling-sound profile spanning 440 Hz – 4.7 kHz, 440 Hz – 13.1 kHz, or 440 Hz – 8.9 kHz; crevice collisions triggered additional "knocking" bursts to heighten realism. Average tactile forces increased systematically with cue intensity: 0.40 N, 0.79 N and 0.88 N for visual-only trials and 0.41 N, 0.81 N and 0.90 N for audio-only trials on Terrains 1-3, respectively. Higher audio frequencies and denser visual textures both elicited stronger muscle activation, and their combination further reduced the force needed to perceive surface changes, confirming multisensory integration. These results demonstrate that consumer-grade isometric devices can reliably induce, and measure graded pseudo-haptic feedback without specialized actuators, opening a path toward affordable rehabilitation tools, training simulators and assistive interfaces.

## 1    Introduction

Haptics—the scientific study of kinaesthetic [1] and cutaneous sensation—has transformed human–computer interaction by letting users "feel" virtual objects through force and vibration cues [2]. Yet the specialized motors and actuators needed for true force-feedback remain bulky and prohibitively expensive, restricting widespread adoption outside research labs and high-end simulators. By shifting the burden of "touch" [3] from hardware to multisensory integration, pseudo-haptic techniques promise low-cost, software-defined haptic experiences that scale to commodity phones and tablets. Such communicative technology lets us evaluate the relation between human sensing and its possible applications in different domains like robotics, psychophysical studies, and bionics [4]. An emerging workaround is pseudo-haptics [5] perceptual illusions in which the brain infers forces, textures or weight purely from carefully synchronised visual or auditory signals, with no mechanical actuation. A scion of haptic technology which is associated with studying the illusion of haptic or false force generated within human physiology when the person is subjected to sensory cues [6]. Despite two decades of proof-of-concept demonstrations in HCI [7], we still lack quantitative evidence on how specific visual and audio parameters [8] jointly modulate perceived pressure, and whether consumer-grade isometric devices can both induce and measure such effects with sufficient fidelity for rigorous psychophysics [9]. Addressing this gap, we introduce a Unity-based Rollball platform deployed on a 10-inch tablet and instrumented with a Robotous RFT40 force–torque sensor. Participants roll a virtual ball across three textured terrains while terrain-linked rolling sounds (440 Hz–4.7 kHz, 8.9 kHz, 13.1 kHz) provide graded auditory cues; crevice collisions trigger brief "knock" bursts to heighten realism. Continuous fingertip force and touch-radius telemetry allow milli-newton-level tracking of motor responses in real time.

Our study makes four contributions:

**Affordable multimodal testbed:** We demonstrate that a £200 tablet plus an off-the-shelf force sensor can replicate key aspects of high-end haptic rigs, enabling rapid, portable experimentation.

**Systematic cue–force mapping:** By decoupling audio and visual channels, we show that higher spatial-frequency textures and higher-pitch sounds each elicit stronger voluntary pressure [9], and that their combination produces super-additive effects, confirming multisensory integration.

**Psychophysical benchmarks:** We report just-noticeable differences (JNDs) for surface changes below 0.05 N and establish baseline pressure ranges (0.40–0.90 N) that future pseudo-haptic interfaces can target.

**Application outlook:** The findings open pathways toward inexpensive motor-rehabilitation trainers, tactile accessibility tools, and immersive gaming where software alone sculpts convincing touch illusions.

By quantifying how visual–auditory cue design translates into measurable motor output, this work advances pseudo-haptics from intriguing illusion to engineering principle, and provides an open, low-cost platform for the community to build upon.

## 1.1 Pseudo Haptic Feedback

Feedback, defined as an information about reactions to a product and pseudo-haptic feedback is the body's reaction towards the sensory cues (visual and audio cues) engrossing user's tactile senses and allows the electronics devices to establish a communicative medium with the human body. Pseudo haptic feedback can be derived from the isometric devices which in turn hack the somatosensory signal [10] and visual cortex signaling [11] to give rise to an illusion of communication between the virtual object and our body. Predominantly, visual cues and audio cues are the most effective signaling methods to generate pseudo-haptic sensations. Tactile sensations encompass a wide variety of sensations (not only vibration or pressure) but sometimes subjects can also sense pain, temperature, and apparent position change [12]. Physiologically, psycho-physical actions [13] like muscle motion allow subjects to exert muscle force even in the absence of any object.

## 1.2 Unity Engine Application

Unity engine [14], a versatile designing platform used for creating both 2D and 3D interactive applications that allows developers to build C# programming language-based applications. It is known for its user-friendly interface, extensive asset library, and real-time rendering capabilities, making it ideal for game developers across various platforms, including android and windows. With built-in support for physics (motion, friction, drag, force, gravity), lighting, and animation, unity simplifies the process of game development while allowing for complex, dynamic interactions within a virtual environment. The platform's ability to easily deploy games to multiple platforms is a significant advantage in reaching broader audiences.

For this research an android based unity3d gaming application named "Rollball" is designed. It is an easy to play virtual game where three terrains (each having different textures) embedded with three different sound frequencies are created and then a ball (obeying physics laws of motion) is allowed to roll over these terrains. The game was allowed to be played on the tablet where the single point pseudo haptic interaction was established between the game and player. The maximum speed of the ball was hard coded for all three surfaces. The occurrences of hinderances in the form of holes and crevices were created in Terrain 2 and Terrain 3 which forces the player to apply more pressure to the touchscreen of the tablet. Terrain 1 (a plain surface, providing no obstruction to the ball) was identified as the baseline system for our model as it required minimum amount of force to roll the ball over it to evaluate the visual and auditory cues [15] and to understanding their importance in they generate the pseudo haptic sensations in users. Additionally, we added the three audios (in .mp3 format) to the to the terrains and allowed application players to roll the ball over the three surfaces. Multiple experiments were conducted, first with visual cue only, second with auditory cues only and third with both visual and auditory cues and special short burst of noise was also added to the scene for Terrain 2 and Terrain 3 to be played when ball gets stuck in the holes or crevices.

## 1.3 Robotous Sensor

To quantify the force applied by the players over the touch pad (touch pressure readings/direct touch or stylus pressure reading/indirect touch) we have used the RFT series (RFT40-SA01-D) force torque sensor. It can provide the real time tensile and compressive force applied by the players/users. It is a capacitive based sensor with 40(D)x18.5(H) mm | 60g dimensions with an overload capacity of 100N.

## 2 Literature Survey and Related Works

### 2.1 Visual Cue Studies.

Multiple studies have been conducted over the visual cues and how the human brain reacts to different visual cues. Yoshida et al. 1968 used the SD method, where 20 pairs were rated as they explored the materials with vision by letting them touch the materials [16] Lyne et al.1984 [17] experimented with hardness and roughness using 8 types of tissues and paper towels using similarity estimation and MDS. Hollins et al. (2000) [18] used 17 textures materials. Parallelly, Tamura et al. [19] uses SD method using 15 different materials (rubber, wood etc.) for 20 participants and constructed three dimensions. In 2003 and 2004 Picard et al. experimented with 20 participants to classify 24 material using similarity basis estimation [20]. Gescheider et al. identified visual dimensions of protruded textures and Pacinian channels were employed [21]. Here plasticized dots of trapezoid shape were created having an inter dot spacing from 1.34 mm to 5.93 mm. Guest et al. in 2011 [22] used factor analysis to analyze the tactile dimensions of fabrics cotton, latex, hessian, and silky and polyester with an aim to find the location on the body contacting the surface effects the tactile dimensionality. Min Li et al. in 2016 proposed a CAD based simulation to convey simulated soft surface simulation stiffness information [23] via comparing multi-point pseudo haptic interaction with the single point-pseudo haptic interaction through creating soft surface and moving a simulated avatar over it.

Swamp-experiment, based on a study of Lecuyer [24], also answers the principle of pseudo haptic feedback in virtual environments. Users were asked to move a virtual cube in an Insilco environment having a constant velocity in a particular direction and cube velocity was controlled by a force sensitive device.

### 2.2 Auditory Cue Studies.

*Coalescing Sensory Cues:* Human physiology is a complex system that acts one-as-a-whole. Here information processing takes place on multiple levels, visual and audio inputs to the human brain connect the somatosensory and visual cortex and both processes the received cues to generate the reaction (psychophysical) towards the input. Temporal integration of multiple cues is very important in understanding the digital setting. Users often combine the inputs from multisensory modalities depending upon their temporal proximity [25]. Time oddities between sensory cues could result in biased estimation about object (virtual or real) characteristics [26]. Auditory cues can be counted as reliable for generating haptic sensations with the user [27].

*Optical Instigations:* Studies have shown effects of optical cues on the psychophysical domain of the participants, achieved by seeing the changes in object displacement and where delays are manipulated. The bigger the displacement, the greater the amount of sensation generated [28]. Delay in visual feedback is quite important pseudo sensations generating cause. In Honda et al. [29], the relation between motion of arm and the motion of a virtual object was identified and a delayed based differential was established in mass. Such pseudo sensations are because the mechanical load model within the brain gets reselected [30]. Summers et al. [31] used 10 diverse papers as banknotes to create a perceptual space when they were allowed to touch them using both hands.

*Delay Visualization of Auditory Cues*: Audio feedback was employed with some delay with respect to the click of the mouse [31]. When multiple ranging was experimented with, causality and proximity proved a close relationship to each other. The feedback here allows the users to generate a good amount (greater than average) of muscle response to virtual object motion.

*Sound Effect based Pseudo-Haptic*: "Parchment Skin Illusion" means any change in auditory feedback can induce a change in texture [32]. Any delay to auditory feedback weakens the effect of pseudo feedback [33] [34] suggesting that the time domain regularity between an action of a person and sound provided is necessary for generating a pseudo haptic effect.

*Sound Based Identification of Materials*: Authors here manipulated the delay in the auditory feedback along with the frequency and loudness of the sound effect. When a low-pitched sound effect was given when the user was holding a small paper box, they rated the weight of the box as heavier in comparison to the high-pitched sound effect [35]. Another research examining the effect of auditory feedback on performance when picking a virtual object found that user interpreted the object heavier in comparison to when the sound pitch played during the object picking was lower [36]. With such a development in technology, there are almost endless possibilities where pseudo-haptics can be implemented. Team is also looking into the potential of this technology into the Genomics and Epigenomic studies [37], Neurodegenerative disorders and their associated surgical intervention [38] and studying Deep Learning via Spiking Neural Networks and understanding human cortical system better [39].

## 3    Methodology

### 3.1    Demographic Profiles (Experimental and Control Group)

For experiments 4 individuals (2 males and 2 females) were identified and selected having an age between 22 to 26 years to play the designed game. The chosen candidates were extensively interviewed on diverse topics like previous video game experience and their medical histories. As the research is associated with manipulating users' experience with visual cues, we intentionally avoided candidates having any medical history of seizures, migraine, or any neurological disorders to avoid any mishap during these experimentations. All 4 candidates have had some experience in playing video games earlier. Audio-only candidates relied only on sound signals variations to interpret the environment, and by removing visual feedback he faced a little problem in understanding the terrain of friction and resistance. During the experiment players were almost always found to have applied a little more pressure over the tablet than the visual candidate. Each time when ball is stuck player muscle tension changes significantly and she/he tries to force his way out of the hole (even knowing that the maximum drag that can be applied to the ball is hard coded) hence showed less accurate and control or response. On the other hand, visual only candidates' responses were based on only visual cues (terrain textures, topography, obstructions, and location of holes). These candidates are found to apply less pressure in all three terrains, even when the ball is stuck at the crevices compared to the audio-only candidate.

Some of the prime differences between the experimental group candidates were that the audio-only player mostly tried to escape the hole in the forward direction and the video-only player tried all other directions as well (back and sideways). The ocular movement is greater in visual only candidates. Auditory candidate presented a greater muscle tension, not just in finger and hand muscle but also in body. When both auditory and visual cues are combined, the participants have access to more sensory input, providing the richest form of interaction. This condition helps evaluate whether combining these sensory modalities strengthens or alters the perception of haptic feedback, simulating a more realistic interaction with the digital environment.

Control group consisted of 2 candidates (1 male of 34 years of age and 1 female of 33 years). Both candidates have a history of experience with video games and field sports. Candidates were given limited auditory and visual cues. Their interaction with the touch was restricted to 1 second taps to direct the ball over the terrains to establish a baseline. Both players struggle to understand the key dynamics and mechanics of the game such as terrain change, friction change, and ball speed which leads to less accurate and even less immersive game play.

In summary, the control group with restricted cues would experience more challenges compared to the experimental groups, highlighting the importance of sensory feedback in pseudo-haptic studies. The experimental groups demonstrated superior responses as they benefit from auditory, visual, or combined cues to interpret and interact with the game environment.

## 3.2 Hardware Setup- M10 HD Lenovo Tablet

Lenovo M10 HD 2nd Gen 10.1" tablet was utilized to execute the Android (Operating System) application. The tablet Operating System was equipped with auto calibration of touchscreen when switched off and brought back online. To identify the optimum conditions of experiments, the tablets' touch capabilities were kept at pristine conditions to avoid faulty readings.

## 3.3 Sensory Cues

*Sensory signals* are derived from the sensory input provided by a receiver. These cues are then analyzed and measured to understand the properties of the surrounding environment as perceived by an individual. Cues extrapolate meaningful data from the information received. Examples of sensory cues include visual, auditory, and haptic cues. For this investigation, we focused on using visual and auditory cues to elicit pseudo-haptic feedback among the participants. Visual cues are captured by the eye and processed by the human brain's visual cortex system. They provide significant information about an individual's three-dimensional spatial surroundings. Similarly, auditory cues are received through the ear and processed by the auditory system. Audio feedback can trigger somatosensory responses.

We created visual cues as part of a 3D virtual gaming application that the user could interact with. These visual cues are textured surfaces called terrain. Terrain1, a plain surface having a smooth, unobstructive visual experience. Terrain 2 appeared textured equivalent to the gravelly-spiky topography with gentle and uniform slopes, while Terrain 3 resembling mountainous topography having steeper and varied slopes.

For auditory effects gaming application was embedded with three sound effects having frequency ranges from 440 Hz to 4.7 KHz, 440 Hz to 13.1 KHz (knocking sound) and 440 Hz to 8.9 KHz (Fig.1). The selection of these specific audio frequencies was a deliberate choice, aimed at maintaining a high level of realism in the experiments. Extensive auditory testing with various games, videos, and audio samples was conducted, ultimately leading to the identification of these three frequency ranges as the most suitable for creating an authentic experience of ball rolling over smooth and textured surfaces and signaling. Once selected, the frequencies were analyzed through Python programming (jupyter notebook) to identify their range as shown in Fig.1.

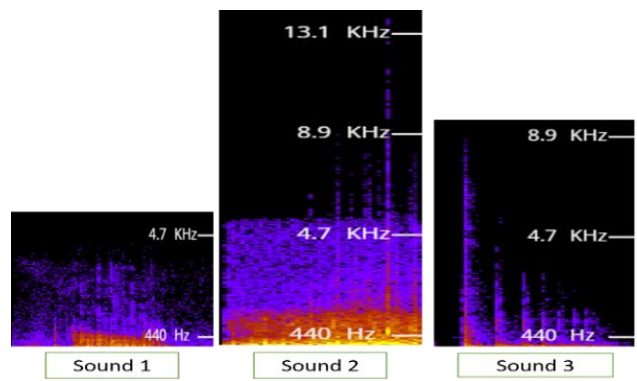

**Fig. 1.** Frequency Spectrum of Sound Effects

A continuous short burst of noise resembling a knocking sound (Sound 2) was introduced into Terrain 2 and Terrain 3 when the ball encounters an obstruction and gets stuck in holes or crevices. The purpose of this knocking noise was to enhance the pseudo-haptic feedback by simulating a sensation of ball getting stuck or experiencing resistance, promoting the player to apply more pressure on the touchscreen.

The sole intention of these embedded frequencies was to provide distinct auditory experiences corresponding to the terrains, potentially affecting the players' perception of ball movements and the texture of the surfaces.

*Controlling and Varying Stimuli:* The sound frequencies were programmatically controlled. When the ball rolled over the surface with no obstruction (Terrain 1), only sound 1 (smooth ball rolling sound) was heard. When the ball rolled over Terrain 2 and Terrain 3, a different sound (sound 3) was heard, along with a knocking sound (sound 2) when the ball lost its motion. Hence, the sound frequencies and visual textures associated with the terrains were consistent/hardcoded, ensuring each player experienced similar visual and auditory cues when interacting with a specific terrain. Therefore, sound frequencies and visualizations (terrain textures, holes, and crevices) were strategically placed to challenge the user's perception and influence pseudo-haptic feedback.

### 3.4 Unity Software Application

While creating a Rollball game, unity's robust terrain design tools and physics engine played a crucial role. Project involved designing three distinct surfaces, each with unique textures and varying resistances to motion, adding complexity and challenge to the game. Unity's terrain editor was used to sculpt these landscapes, allowing for customization in terms of elevation, texture, and smoothness. By adjusting the terrain's texture and friction properties, the rolling ball experiences drag and resistance as it moves across each terrain. Designing the metallic ball was equally important, and it involved creating a simple 3D sphere with materials applied to achieve realistic texture and appearance. The ball's motion adhered to physical laws we owe many thanks to unity physics engine and rigid body component which governs interactions like gravity, friction, and drag. These features are fundamental in maintaining the realism of the rolling motion. Gravity, which pulls the ball downwards, was adjusted to simulate a real-world environment, ensuring that the ball rolls in response to player input (keyboard and touch controller) while adhering to the boundaries of each terrain.

A significant technical component of this project was designed using a unity C# programming language such as ball control, touchpad visuals, and inputs. In unity, C# scripts allow for precise control over game mechanics, such as how the ball responds to user inputs. For touchpad devices, such as tablets and smartphones, Unity offers touch input APIs to detect user interactions. These APIs handle single touch events, allowing the player to roll the ball using direct touch. This application, compatible with both Android and Windows operating systems, enables the player to guide the ball across various textured surfaces (Fig. 2) through either direct or indirect touch. The game was developed to comply with fundamental physical principles, including friction, drag, and gravity. To establish a better baseline, we set a maximum speed limit for the ball's rotation.

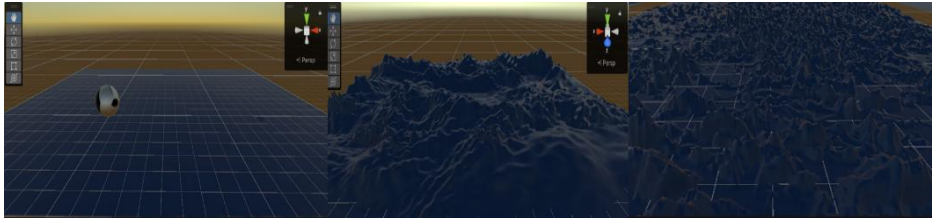

**Fig. 2.** Terrains
(a: Terrain 1 (Left), b: Terrain 2 (Right), c: Terrain 3 (middle))

Surface1 (Fig. 2a), a regular plain surface offers the least resistance to the ball when rolled over it. This surface requires minimal force to move the ball, allowing the user to apply minimum tactile force on the tablet. When provided only visual cue, the average force applied to move the ball noted was 40.62 grams (0.39834 N). With only auditory cue this changed to 41.35 g (0.40550 N). Terrain 2 (Fig. 2b) was designed to mimic a gravelly, uneven surface with holes and crevices (Fig. 4), causing the ball to get stuck at times. This was done to study the pressure users applied to roll the ball out of these obstacles, even though the speed couldn't exceed the maximum set limit. On this surface, users applied the average pressure of 80.45 grams (0.7889 N) (which is twice the surface 1) when given only visual cue and a force of 82.37 grams (0.8077 N) when only auditory cue was given. In the areas where the ball became stuck, users applied pressures of 100 grams or more (0.9806 N). This additional force applied to the touchpad demonstrates the psychophysical phenomenon of users feeling compelled to apply extra pressure to free the ball from the hole.

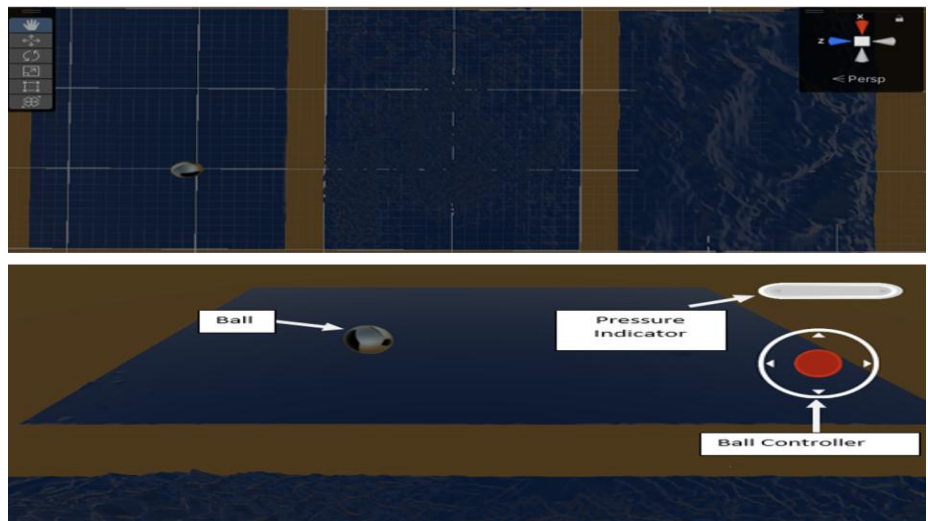

**Fig. 3.** Terrains Top View (Top) and Console View (Bottom)

To safeguard our equipment, we restricted the uninterrupted pressure applied to the tablet's touchscreen to a duration of 3 seconds. Surface 3 (Figure 3c) features a gray, mountainous texture. On this terrain, the ball was able to ascend and descend minor cliffs, and corresponding data was logged. The ball effortlessly traversed the base of these elevated features. A minimal pressure of only 50±2 grams (0.49033 N) was observed in conjunction with both visual and auditory prompts. To navigate the cliff ascent, the user needed to apply a force of 90 grams (0.8825 N) or greater when visual cues were present, while an auditory cue prompted a pressure of 92 grams (0.9022N).

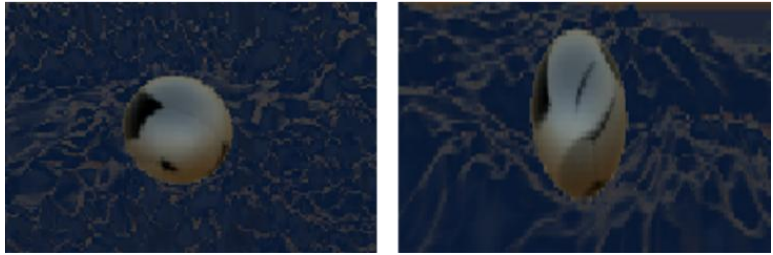

**Fig. 4.** Crevices and Holes in Terrain 2 (left) and Terrain 3 (right).

Another method, which we consider a potentially novel approach, was adopted to quantify the pressure applied to move the ball by providing visual cues such as pseudo-haptic feedback. This was achieved by measuring the pressure generated by a touch on the screen of the isometric device. When the player touches or taps the screen, a "touch effect circle" appears around the point of contact. The radius of this circle, known as the touch radius (as shown in Fig. 5), is directly proportional to the pressure applied by the player to move the ball. The more pressure applied, the larger the touch effect circle, and the higher the touch radius value. This radius is used to calculate pressure levels and is mapped to a Pressure Indicator, providing feedback to the player.

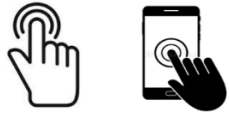

**Fig. 5.** Touch Circle Radius

### 3.5 Robotous Sensor Readings

The RFT Series Force torque sensor (Figure 6) was utilized to gauge the tactile force exerted by either a finger or a stylus on the tablet, which guided the ball within the simulated environment. This sensor operates capacitively, precisely measuring tactile and compressive pressure, calibrated in Newtons. Its capability extends to accurately measuring forces up to 100 Newtons. While typically integrated into robotic arms, this sensor proved to be an invaluable component for our project. It facilitated the quantification and verification of the relationship between the feedback provided to users—which was dependent on textured cues—and the force they applied to the simulated virtual object, specifically a Lenovo tablet. Furthermore, by analyzing the compressive force readings, we can confidently affirm the presence of psychophysical variations as visual and auditory cues evolved over time. Given that this application is designed for single-point contact, only touch pressure data from a single finger or stylus was recorded for this research.

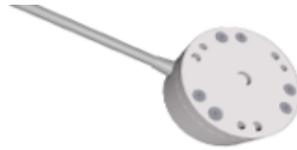

**Fig. 6.** Force Torque (Robotous) Sensor

### 3.6    Hardware Calibration

To ensure consistency, reliability, and reproducibility of results across the different experiments, various measures were implemented. All experiments were conducted in a controlled environment, avoiding sudden changes in temperature and humidity to maintain stable conditions. Excessive force or torque on the sensor was avoided to prevent damage and inaccurate readings. The Lenovo tablet was equipped with a toughened, scratch-resistant screen protector to prevent scratches and damage that could affect sensitivity over time. Post-experiment feedback from users was also collected to identify and address any touch performance issues that may have arisen during the experiment, thus ensuring long-term reliability.

## 4    Experiments and Discussion of Outcomes

Below we are evaluating the research outputs, our research project was divided within many fields of study out of which we are focusing on presenting the gaming application's review, text on audio tactile relation (estimate of relationship between an audio cue, visual cue and the pressure data exerted over the tablet) and at the last we have added a comment over the findings of the psychophysical analysis.

### 4.1    Experimental Setup (Hardware and Software)

*Hardware setup: Isometric device and Robotous sensor.*
We utilized a Lenovo M10 HD 2nd Gen 10.1" tablet, which features integrated touch-sensing capability. Our system integrated the robust RFT series force-torque sensor to quantify the pressure differential applied by the player when guiding the ball across three distinct textured surfaces, positioned atop the tablet's sensor.

*Software Setup: Unity Application*
The gaming application played multiple times, and for each instance, the tactile force exerted on the tablet—whether through direct (finger) or indirect (stylus) touch—was recorded. Table 1 below presents the average tactile pressure applied over each terrain, as measured by the Robotous sensor. We observed that the applied pressure directly correlates with the texture density; consequently, denser textures or more pronounced visual/auditory cues resulted in greater pressure activity variations within the user's finger muscles.

When a user taps the device screen with their finger, it produces a "Touch Effect Circle" that appears around the point of contact. The radius of this circle is directly proportional to the pressure applied on the screen. Therefore, as more pressure is applied, the touch effect circle expands, and its radius value increases. Pressure intensity (Fig. 7) is represented by color on the Unity display. The initial pressure is indicated by a green color, and as continuous pressure is maintained over the touch screen, the color gradually shifts to red, signifying an increase in the amount of pressure being applied on the touchpad.

$$P = T_e \times t_d \qquad (1)$$

where $T_e$ is the touch even or touch effect circle and $t_d$ total duration of touch (in seconds).

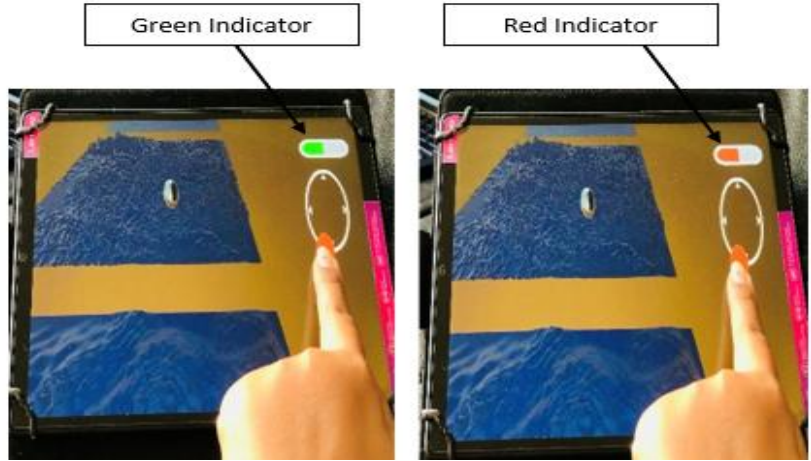

**Fig. 7.** Game Play Application

## 4.2 Audio Tactile Correlation

The audio signal variations experienced across different terrains stimulated muscle (in finger) contractions in the user/player. These contractions prompted the player to exert progressively more pressure as the terrain changed (and the audio frequency increased). Consequently, the pressure output directly correlated with the rise in audio frequency. Although the visual cues within the application also significantly contributed to inducing pseudo-sensation, our research further validated the direct connection between audio cues and tactile responses. Below, we present a table detailing the average pressure (from 10 touches/taps) across the various terrains. Given that Terrain 1 is simplistic and lacks associated texture, the audio signal played continuously without interruption, resulting in less force being applied. Terrain 2 features a dense texture, and our audio was engineered to produce a pulse noise each time it encountered an obstruction, leading to the highest force/pressure exerted on the touchpad. Terrain 3, in turn, possesses a texture density greater than Terrain 1 but less than Terrain 3. Therefore, the collision sound frequency is higher than Terrain 1's but lower than Terrain 3's, yielding an intermediate pressure value.

| Terrain | Mean Pressure Applied For Visual Cues (In Newtons) | Mean Pressure Applied For Auditory Cues (In Newtons) |
|---------|------------------------|------------------------|
| Terrain 1 | 0.39834 N | 0.40550 N |
| Terrain 2 | 0.7889 N | 0.8077 N |
| Terrain 3 | 0.8825 N | 0.9022 N |

Table.1. Average Tactile Pressure

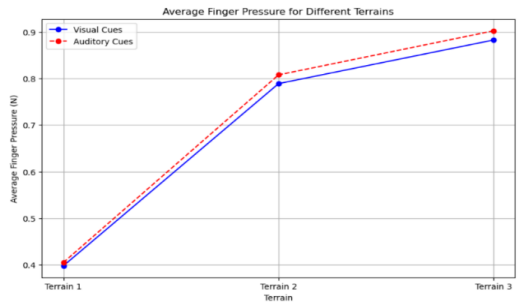

Fig.8. Plot for Table 1 (Average Tactile Pressure)

*Touch Radius*

The radius values produced by pressure applied to the Lenovo Tablet screen ranged from a minimum of 13 to a maximum of 57 pixels. These measurements are expressed in pixel units. It is important to note that these values are device-dependent and do not remain constant based on the size of the input method used (in our study, either a finger or a stylus) (Table 2)

| Terrain | Pixel Range | Indicator Values (0-10) |
|---------|-------------|--------------------------|
| Terrain 1 | 13-34 | 0.69 – 2.82 |
| Terrain 2 | 19-58 | 0.97 – 6.15 |
| Terrain 3 | 13-56 | 0.97 – 6.14 |

Table.2. Touch Radius Readings (For Visual Cues)

### 4.3 Pressure Trend Analysis

The provided plot (Fig.8) represents the increase in pressure when user/player relies solely on auditory cues alone. This difference between visual and auditory modalities is not significantly large. There is still an established consistency among the two cues suggesting the fact that human physiology can adapt well. Further study with better sensors and haptic feedback devices can potentially corroborate the readings and provide better results.

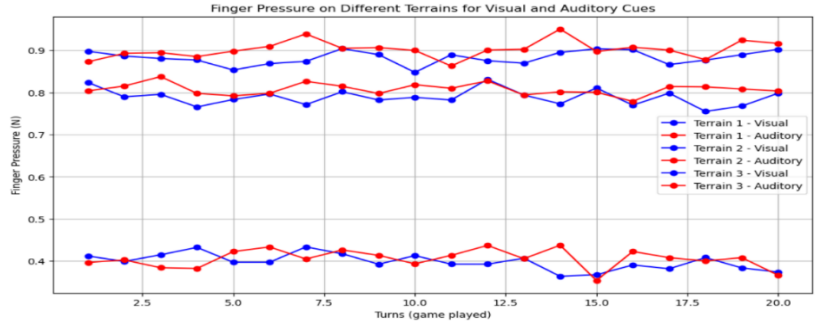

Fig. 9. Finger Pressure Trend Analysis

### 4.4 Psychophysical Analysis

This analysis is linked to the tension generated within the users' finger muscles as they maneuver the ball across various terrains in the Unity gaming platform. Rather than employing muscle sensors, we utilized input provided directly by each player. Users were also requested to note what they perceived as the "Just Noticeable Difference (JND)" or "Difference Threshold," which is defined as the minimum level of stimulation an individual can detect over 50% of the time. It can be computed as:

$$K = \Delta I \ / \ I \qquad (2)$$

Here $\Delta I$ represents the introduction of additional stimulation beyond the original intensity, $I$ denotes the original intensity of the sensation, and K is the weber constant.

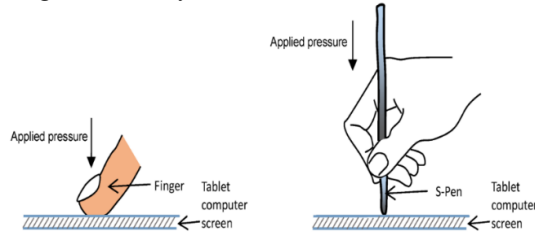

Fig. 9. Concept Art for User Haptic Interaction

Users reported perceiving a difference in the required force for touchpad interaction almost immediately when the terrain or auditory feedback changed. A comparable amount of tactile force was implied, and similar JND feedback was acquired from the users.

Due to the inherent time constraint of two and a half months, this study primarily aimed to establish a proof-of-concept for assessing perceived pseudo-haptic sensations. While psychophysical measures such as skin conductance and heart rate were not employed in this phase, their inclusion in subsequent research would undoubtedly offer valuable corroborating evidence for the observed phenomena.

### 4.5 Ethical Considerations

While the integration of pseudo-haptic sensations offers significant advancements in user experience and interaction, it is crucial to address the ethical implications, particularly "Informed Consent" where participants must be fully informed that their sensory experience will be manipulated artificially. "User Autonomy", as sensory experiences might impact users' ability to make fully autonomous decisions. "Dependence and Overreliance", where users might become overly reliant on induced sensations.

This research is fully compliant with the General Data Protection and Regulation (GDPR) guidelines where the users/players were fully informed about the research and how their data will be used and processed. All the participants were informed about the procedures and the nature of the data collected and the option of withdrawing from the study. Only the data which is strictly required was collected and the personal data collected was anonymized through unique codes to protect the participants' identities. The collected data was stored in a secure environment with access restricted to authorized personnel only. And no immediate or identifiable future risks were identified here.

## 5 Conclusion

This study successfully produced, assessed, and quantified the pseudo-haptic sensation, which was simulated on a Lenovo tablet (an Isometric device) as an Android application, incorporating both visual and auditory feedback. The findings include tactile pressure measurements recorded on the tablet via a Robotous sensor, alongside muscle reactions (quantified by the noticeable difference). We achieved the anticipated outcomes concerning the Texture Pressure analysis, where users exhibited distinct psychophysical responses. These artificially induced sensations manifested as muscle tension, which ultimately resulted in physical pressure being exerted through finger contact with the tablet. Furthermore, a positive audio-tactile correlation (involving frequency variations) was established. This research also opens avenues for the algorithmic quantification of tactile force based solely on finger-pixel interaction.

The overall greater pressure was implemented when audio only cues were provided and the average tactile pressure recorded over Robotous sensor for visual-only experiment group are 0.39834 N, 0.7889 N, 0.8825 N which are lesser than the average tactile pressure recorded for audio-only group (0.40550 N, 0.8077 N, 0.9022 N) with greater muscle tension recorded in finger, arm and body for participants only provided with audio frequencies.

We can also corroborate the positive generation of Pseudo-haptic sensations using only audio signals within the arm of the user/players and different auditory frequencies varying the effects of Pseudo-haptics. The study correlates with visually generated sensations. Users documented significant tension in finger and arm muscles when extra force applied by the users (when ball was stuck in holes and crevices).

An intriguing observation from this investigation is that subjects exhibited a significantly higher application of force when provided with exclusively auditory feedback compared to a condition utilizing solely visual feedback. This finding, along with the potential utility of commercially available, low-cost instrumentation for research in this domain, strongly suggests the presence of sensory compensation and an impaired calibration capacity. We hypothesize that this phenomenon is the underlying mechanism contributing to diminished feedback control and altered perceptual processing.

## 6    Future Works

Regarding future research directions, audio-induced pseudo-haptic sensations warrant a more extensive investigation. Further analysis could involve examining the varying pitches, amplitudes, and other characteristics of auditory signals. Such exploration holds significant promise for the creation of applications beneficial to psychological studies, as well as the development of biomedical assistive technologies for amputees and individuals with physical disabilities. This type of technology could also be highly impactful in the field of biomedical robotics, offering substantial aid to medical practitioners and surgeons in remote operations. Further studies are needed in collaboration with Neuroscientists employing Electromyogram and Electrooculogram to understand altered perception.

One of the potential applications with the improvement of Pseudo-Haptic we have in mind is to pursue Therapeutic Intervention for Burn Victims. We believe Pseudo-haptic can play a pivotal role in Distraction Therapy research for immediate pain management; it surely has the potential to help patients engage in virtual environments that are soothing, cold and/or distractive. Another area of research could be Sensory Reeducation for patients who have undergone skin grafts or may have altered sensory perceptions caused by burns.

Visual and auditory cues are instrumental and valuable for creating pseudo-haptic sensations, but they have some limitations in terms of tactile realism and precision in sensory responses. Auditory cues can convey spatial information to a limited extent and are very few effective in providing detailed information about the environment. We can surely explore additional modalities to overcome the above documented limitations, such as use of wearable devices that provide proprioception, using exoskeletons to enhance kinesthetic interactions, there is also a possibility of integrating thermal and airflow feedback to simulate temperature change and mimic the sensation of object moving past the skin.

The team also plans to integrate this research with Ultraleap's ultrasound-based touch array and the MediaPipe framework, aiming to explore the fusion of computer vision (hand tracking) technology with haptics to enhance and manipulate tactile sensations.

# 7 Acknowledgement

We extend our profound thanks and sincere gratitude to our colleagues and academic mentors for their unwavering assistance and direction across all phases of our research endeavors. I wish to convey my deep appreciation to Professor Dr. Kaspar Althoefer, who leads the distinguished Centre for Advanced Robotics @ Queen Mary (ARQ), for furnishing crucial technical support and providing essential encouragement throughout this project.

In addition, we acknowledge that this research builds upon our own prior MSc dissertations conducted at Queen Mary University of London: Pseudo Haptic Study: Its Medical and Prosthetic Application [40] and Pseudo Haptic Feedback Study for Isometric Devices [41]. These works provided the experimental and conceptual foundation for the present study, particularly in the exploration of pseudo-haptic feedback through auditory and visual modalities, and the application of the Robotous RFT sensor and touch radius analysis in quantifying user-applied pressure.

# 8 Personal Acknowledgement

Lastly, I fondly recall my beloved dog, "Disney, the Pug", whose companionship and steadfast presence offered both solace and inspiration during this journey. She is deeply missed.